\documentclass[10pt,conference]{IEEEtran}

\usepackage{amsmath,amsfonts}
\usepackage{algorithmic}
\usepackage{graphicx}
\usepackage{textcomp}
\usepackage{listings}
\usepackage{xcolor}
\usepackage{cite}
\usepackage{csquotes}
\usepackage{xurl}
\usepackage{subcaption}
\usepackage{hyperref}
\usepackage{comment}
\usepackage{makecell}
\usepackage{todonotes}
\usepackage{subcaption}
\usepackage{booktabs}
\usepackage{glossaries}
\usepackage{url}
\newglossaryentry{LLM}
{
  name={LLM},
  description={Large Language Model},
  first={Large Language Model (LLM)},
  plural={LLMs},
  descriptionplural={Large Language Models},
  firstplural={Large Language Models (LLMs)}
}

\newacronym{ML}{ML}{Machine Learning}
\newacronym{GenAI}{GenAI}{Generative Artificial Intelligence}
\newacronym{AI}{AI}{Artificial Intelligence}
\newacronym{MoE}{MoE}{Mixture of Experts}
\newacronym{NVP}{NVP}{N-Version Programming}
\newacronym{KL}{KL}{Knight-Leveson}
\newacronym{CoT}{CoT}{Chain-of-Thought}
\newacronym{ICL}{ICL}{In-Context Learning}
\newacronym{AST}{AST}{Abstract Syntax Tree}
\usepackage{balance}

\usepackage{float}
\newfloat{listing}{tbp}{lol}
\floatname{listing}{Listing}

\lstdefinestyle{smallc}{
    language=C,
    basicstyle=\ttfamily\fontsize{6pt}{6pt}\selectfont,
    backgroundcolor=\color{gray!10},
    frame=single,
    rulecolor=\color{gray},
    breaklines=true,
    showstringspaces=false,
    tabsize=2,
    captionpos=b,
    numbers=left,
    numberstyle=\tiny\color{gray},
    numbersep=5pt,
    xleftmargin=5pt,  % Indent slightly from the left
    xrightmargin=5pt
}

\lstdefinestyle{smallcpp}{
    language=C++,
    basicstyle=\ttfamily\fontsize{6pt}{6pt}\selectfont,
    backgroundcolor=\color{gray!10},
    frame=single,
    rulecolor=\color{gray},
    breaklines=true,
    showstringspaces=false,
    tabsize=2,
    captionpos=b,
    numbers=left,
    numberstyle=\tiny\color{gray},
    numbersep=5pt,
    xleftmargin=5pt,  % Indent slightly from the left
    xrightmargin=5pt
}

\lstdefinestyle{smalljava}{
    language=Java,
    basicstyle=\ttfamily\fontsize{6pt}{6pt}\selectfont,
    backgroundcolor=\color{gray!10},
    frame=single,
    rulecolor=\color{gray},
    breaklines=true,
    showstringspaces=false,
    tabsize=2,
    captionpos=b,
    numbers=left,
    numberstyle=\tiny\color{gray},
    numbersep=5pt,
    xleftmargin=5pt,  % Indent slightly from the left
    xrightmargin=5pt
}

\lstdefinestyle{smallpy}{
    language=Python,
    basicstyle=\ttfamily\fontsize{6pt}{6pt}\selectfont,
    backgroundcolor=\color{gray!10},
    frame=single,
    rulecolor=\color{gray},
    breaklines=true,
    showstringspaces=false,
    tabsize=2,
    captionpos=b,
    numbers=left,
    numberstyle=\tiny\color{gray},
    numbersep=5pt,
    xleftmargin=5pt,  % Indent slightly from the left
    xrightmargin=5pt
}

\usepackage{tcolorbox}
\tcbset{
  colback=gray!10,
  colframe=black,
  boxrule=0.5pt,
  arc=2pt,
  left=1pt,
  right=1pt,
  top=1pt,
  bottom=1pt,
  after skip=1pt,
  before skip=1pt
}

\newif\ifshowcomments
\showcommentstrue
\usepackage{enumitem}
\setlist[itemize]{noitemsep, topsep=-1.0pt}
\def\BibTeX{{\rm B\kern-.05em{\sc i\kern-.025em b}\kern-.08em
    T\kern-.1667em\lower.7ex\hbox{E}\kern-.125emX}}
\begin{document}

\title{A Systematic Methodology for Evaluating Failure Independence in LLM-Generated Code}

%\begin{comment}    
\author{% 
    Rodrigo Pato Nogueira\IEEEauthorrefmark{1}\IEEEauthorrefmark{2}, 
    Karthik Pattabiraman\IEEEauthorrefmark{3},
    Marco Vieira\IEEEauthorrefmark{2}, 
    João R. Campos\IEEEauthorrefmark{1}\\[0.8ex] 
    \IEEEauthorblockA{%
        \makebox[\textwidth][c]{% 
            \begin{minipage}[t]{0.33\textwidth}
                \centering 
                \IEEEauthorrefmark{1}\textit{University of Coimbra}, \textit{CISUC/LASI, DEI} \\ Coimbra, Portugal \\ jrcampos@dei.uc.pt 
            \end{minipage}\hfill 
            \begin{minipage}[t]{0.33\textwidth}
                \centering 
                \IEEEauthorrefmark{2}\textit{University of North Carolina at Charlotte} \\ Charlotte, NC, USA \\ rpatodec@charlotte.edu, marco.vieira@charlotte.edu 
            \end{minipage}%
            \begin{minipage}[t]{0.33\textwidth}
                \centering 
                \IEEEauthorrefmark{2}\textit{The University of British Columbia} \\ Canada \\  karthikp@ece.ubc.ca
            \end{minipage}%
        }% 
    }% 
}
%\end{comment}

\maketitle

\begin{abstract}
\gls{NVP} improves software reliability by executing multiple independent implementations and combining outputs, but its adoption is limited by high cost and the assumption of failure independence, which empirical studies have challenged. Recent advances in \glspl{LLM} reduce the cost of generating multiple implementations, shifting focus to whether their failures are independent.
We propose the first systematic methodology to assess failure independence in LLM-generated code and apply it to 224 problems across twelve models, five languages, and three prompting strategies. We analyze both structural and behavioral diversity (i.e., whether implementations fail on the same test cases), complemented by N-version reliability analysis under majority voting and manual inspection of the generated code. Structural diversity analysis shows that implementations from the same model are highly similar, while different models produce more distinct solutions. The same trend appears in behavioral diversity, with implementations from different models showing higher diversity yet still failing on the same tests far more often than expected under independence. N-version reliability analysis reinforces this: three- and five-version ensembles realize only 0.43 and 0.44 of the reliability gain achievable under independence, dropping below 0.3 when ensembles are built from the same model. Manual fault analysis shows that even different failure patterns often share root causes. Overall, these results suggest LLM-generated solutions do not satisfy NVP's failure independence assumption, though heterogeneous models help partially. They also validate our methodology as a tool for systematically evaluating failure independence as models evolve.
\end{abstract}


\maketitle

\glsresetall

\section{Introduction}

Software faults can lead to severe consequences, ranging from financial losses to safety-critical failures in domains such as aviation and healthcare~\cite{pilkington2024CrowdStrike, wee2024CrowdStrike, perell2020Boeing}. At the same time, faults are inevitable, even in rigorously developed systems~\cite{brooks1987no, adams1984optimizing, ostrand2005predicting}. Fault-tolerance techniques ensure system dependability by preventing unavoidable faults from propagating into observable failures~\cite{chen1978n, laprie1985dependable}.
Among such techniques, \gls{NVP} was proposed as a software analogue to hardware redundancy~\cite{chen1978n}. In \gls{NVP}, multiple functionally equivalent implementations are independently developed from the same specification, and their outputs are combined to mask faults in any single version. The underlying assumption is that independently developed versions are unlikely to fail on the same inputs, thereby enabling fault masking through diversity.

\gls{NVP} relies on the assumption that failures across independently developed versions are independent. An experimental study by Knight and Leveson~\cite{knight2012Experimental} showed that independently developed versions often fail on the same inputs due to shared interpretations of ambiguous specifications, common design strategies, and overlapping reasoning patterns among developers. As a result, the reliability gains are significantly reduced. When combined with the substantial cost of developing and maintaining multiple versions, this has confined \gls{NVP} adoption to niche, high-assurance domains~\cite{Eris2012NVersionRailway, Ron2024NVersionBlockchain}.

\glspl{LLM} are transforming software engineering by automating large parts of development~\cite{chen2021evaluatinglargelanguagemodels, achiam2023gpt, guo2025deepseek}, generating executable code at low cost and scale. Although they remain imperfect on complex tasks~\cite{nogueira2025beyond}, rapid improvements suggest these limitations may be temporary~\cite{wei2022emergent, kaplan2020scaling, li2022competition}. As a result, \glspl{LLM} could mitigate a key barrier of \gls{NVP}: the cost of producing multiple independent implementations. If independently generated versions exhibit sufficiently low failure correlation, \gls{NVP} could become viable with \glspl{LLM}, but this would require methods to assess failure independence and determine whether the reliability gains justify the added cost.

However, no methodology to evaluate failure independence in this context has been proposed. Classical \gls{NVP} studies assume organizational separation of human teams, whereas \gls{LLM}-based generation removes team independence and replaces it with controllable configuration axes (model, language, and prompting); a suitable methodology must therefore measure failure independence at scale while isolating the effects of these factors. At the same time, current research on \gls{LLM} code generation does not address this issue: existing studies primarily focus on functional correctness~\cite{nogueira2025beyond, austin2021programsynthesislargelanguage}, syntactic and semantic diversity~\cite{ron2024galapagos, shypula2025evaluating}, and code quality~\cite{cotroneo2025humanwritten, nogueira2025beyond}, but these dimensions do not reveal whether independently generated implementations fail independently.

In this paper, we propose a systematic methodology for evaluating the failure independence of LLM-generated code, analyzing it across three dimensions: i) structural diversity, measuring how differently implementations are written; ii) behavioral diversity, measuring whether they fail on the same inputs; and iii) reliability gains, measuring how much that failure correlation limits the benefit of combining multiple implementations under majority voting. Together, these three dimensions let us assess diversity at increasing levels of practical relevance, from surface-level code differences to their actual impact on reliability under redundancy. In addition, it includes a manual analysis step to identify whether the underlying faults exhibit common patterns, since failure overlap alone cannot reveal whether seemingly different failures actually share a root cause. We instantiate our methodology in a comprehensive empirical study, evaluating twelve diverse models across 224 problems, five programming languages, and three prompting strategies. This setup lets us draw conclusions from a large, varied set of problems while isolating the effects of model choice, language, and prompting strategy on diversity and reliability gains. The main contributions of this study are: 

\begin{itemize}[leftmargin=*]

    \item \textbf{The first methodology for assessing failure independence in LLM-generated code}, enabling systematic analysis of whether independently generated solutions exhibit correlated failures. Unlike prior work, our methodology goes beyond existing correctness and diversity metrics by directly measuring failure correlation and reliability gains.

    \item \textbf{A comprehensive empirical study} applying this methodology to twelve \glspl{LLM} across 224 problems, five programming languages, and three prompting strategies, resulting in the most comprehensive evaluation of failure correlation in \gls{LLM}-generated code to date. By systematically comparing homogeneous and heterogeneous combinations of models and languages, the study provides evidence of whether deliberately diversifying these choices improves failure independence and reliability gains. The study is accompanied by a publicly available dataset (\href{https://tinyurl.com/CodeDiversity}{tinyurl.com/CodeDiversity}) containing prompts, generated code, unit tests, and metrics, enabling reproducibility and future research.

    \item \textbf{A detailed analysis of structural and behavioral diversity, reliability gains, and fault-level similarity in LLM-generated implementations}, providing insights into the reliability of LLM-generated code under redundancy and helping assess whether multiple implementations improve robustness or share similar failure modes.
    
\end{itemize}

\noindent
Results show the generated code exhibits moderate structural diversity (i.e. how solutions are written). Despite this, solutions often fail on the same inputs, revealing limited failure independence. Combining models reduces failure correlation, but the results remain far from independent. Accordingly, \gls{NVP} provides only modest reliability gains: majority voting increases average reliability from 0.88 to 0.91 when combining five implementations, capturing less than half of the reliability gain that independence would have allowed, with the benefit depending much more on model selection than on programming language or prompting strategy. A deeper fault analysis reveals that seemingly different failure patterns often arise from the same underlying faults, reinforcing the prevalence of shared failure modes. Overall, the proposed methodology provides a practical way to assess whether generated implementations are sufficiently independent for \gls{NVP} and to identify where that independence is lost, supporting informed choices of models, languages, and prompting strategies.

This paper is structured as follows: Section~\ref{sec:background-rw} covers background and related work. Section~\ref{sec:methodology} describes the experimental methodology. Section~\ref{sec:results} presents and discusses results. Section~\ref{sec:ttv} examines threats to validity. Finally, Section~\ref{sec:conclusions} concludes the paper and outlines future directions.

\section{Background and Related Work}
\label{sec:background-rw}

Software fault tolerance encompasses techniques that enable systems to continue operating correctly despite faults. Classic approaches include recovery blocks, which rely on alternative implementations guarded by acceptance tests~\cite{Randell1975RecoveryBlocks}; checkpointing and rollback mechanisms that restore execution to a safe state after failures~\cite{elnozahy2002Rollback}; and structured exception handling~\cite{cristian1995exception}. Design diversity strategies aim to reduce common-mode failures by introducing software heterogeneity. A key example is \gls{NVP}~\cite{chen1978n}, which proposes the independent development of $N$ versions of a software component, each built by a different team, potentially using different languages and algorithms.

The effectiveness of \gls{NVP} depends on the assumption that independently developed versions fail independently. However, Knight and Leveson’s seminal experiment~\cite{knight2012Experimental} showed that such independence rarely holds, as different teams tend to produce implementations that fail on the same inputs. Subsequent studies~\cite{Brilliant1990Analysis} further identified sources of failure correlation, including shared interpretations of ambiguous specifications, similar design strategies, and common practices, highlighting a fundamental limitation of design diversity: independently developed implementations are not necessarily failure-independent. Furthermore, the substantial cost of developing and maintaining multiple versions, combined with these reduced reliability gains, has limited \gls{NVP}'s practical adoption.

The emergence of \glspl{LLM} has enabled automatic code generation from natural language specifications~\cite{chen2021evaluatinglargelanguagemodels, nogueira2025beyond}, reducing the cost of producing multiple implementations. This has motivated work exploring whether LLMs can generate diverse program variants. The Galápagos framework~\cite{ron2024galapagos} showed that \glspl{LLM} can generate diverse variants of a reference program and that such diversity can help mitigate compiler bugs. However, it assumes an initial reference implementation, focuses on a single model, and evaluates diversity in terms of compiler robustness rather than functional correctness or failure independence. Shypula et al.~\cite{shypula2025evaluating} examined functional and structural diversity but not correlated failures, Lee et al.~\cite{lee2025diversely} studied diversity in adversarial robustness, Xue et al.~\cite{xue2024multi} explored multi-version generation focused on API-level diversity, and Zheng et al.~\cite{zheng2025can} measured performance improvement from an N-version ensemble using AST-based similarity between solutions.

Despite these efforts, no work proposes a methodology to evaluate failure independence on \gls{LLM}-generated code, providing an answer to the key question: \textit{do independently generated implementations fail on the same inputs more than chance predicts?} We introduce the first systematic methodology to assess failure independence in LLM-generated code, extending the classical framework with an independence baseline, structural and behavioral diversity metrics, reliability gains, and explicit treatment of the different LLM-based generated configurations axes (model, language, prompt).

%Concurrent with our work, an unpublished preprint~\cite{ron2026nversionprogrammingcodingagents} revisited the classical Knight and Leveson experiment in the context of LLM-generated code. Unlike the original study, it evaluates agentic code-generation systems by replicating the historical experimental setting on a single programming task. While this provides valuable initial evidence that correlated failures remain an important concern in AI-generated software, its objective differs substantially from ours. Rather than reproducing the original experiment, we propose a systematic evaluation methodology for assessing failure independence in LLM-generated code. Our methodology is designed for large-scale black-box evaluation across models, programming languages, prompting strategies, and datasets, and is instantiated on 224 programming problems to study the factors affecting failure independence. While they compare N-version voting only against average single-component reliability, we additionally benchmark against the best individual model and quantify how much of the theoretically achievable gain is realized.

\section{Methodology}
\label{sec:methodology}

\begin{figure}
    \centering
    \includegraphics[width=1.0\linewidth, trim=0pt 8pt 0pt 8pt, clip]{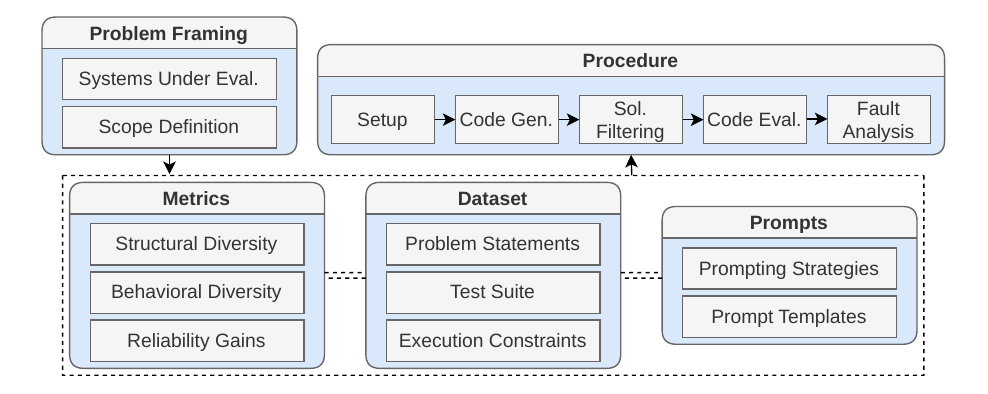}
    \caption{Evaluation Methodology} %\jrc{can we do a more UI-friendly version? :)}}
    \label{fig:framework}
\end{figure}

Figure~\ref{fig:framework} presents the overall structure of our methodology for evaluating whether different \glspl{LLM}-generated solutions exhibit correlated failures. It comprises five components: Problem Framing, Metrics, Dataset, Prompts, and Procedure.

\subsection{Problem Framing}

The problem framing defines the assumptions under which implementations are generated and evaluated, ensuring that observed diversity and correlated failures reflect realistic conditions and that comparisons across systems are meaningful.

\begin{itemize}[leftmargin=*]
    \item \textbf{Systems Under Evaluation.} Characterizing the systems under evaluation requires specifying how \glspl{LLM} are used and what level of access is available. 

    \item \textbf{Scope Definition.} The evaluation scope defines the type of tasks considered. This includes specifying whether models generate complete standalone solutions or partial code, as well as the format in which the input will be given.

\end{itemize}

%\vspace{2pt}
\noindent
In this study, models are treated as black-box systems accessed exclusively through prompts, without modifying internal parameters. This reflects typical deployment scenarios~\cite{brown2020languagemodelsfewshotlearners} and ensures observed similarities or correlated failures result from internal behavior. Models generate complete, standalone implementations from natural-language problem descriptions.

\subsection{Metrics}
\label{sec:methodology-metrics}

Defining appropriate metrics is critical, as they determine how diversity and correlated failures are quantified and interpreted. We consider three complementary dimensions: 

\begin{itemize}[leftmargin=*]

\item \textbf{Structural diversity:} Captures differences in the static structure of implementations, including code organization, control flow, and syntactic constructs, reflecting whether implementations follow different structural approaches. It should be measured among implementations in the same programming language when using language-specific representations; cross-language comparisons require normalization into a language-agnostic representation (e.g., CFGs).

\item \textbf{Behavioral diversity:} Captures differences in execution outcomes across test cases, identifying whether implementations fail on the same inputs more often than expected. It serves as a practical proxy for assessing whether implementations fail for the same reasons, as implementations that fail on the same inputs likely share underlying faults.

\item \textbf{Reliability gains:} Quantify the reliability achieved when combining $N$ implementations under majority voting, translating the degree of failure correlation into a concrete reliability outcome. While behavioral diversity characterizes \emph{how often} failures co-occur relative to an independence baseline, N-version reliability captures \emph{how much} that correlation limits the practical benefit of redundancy.

\end{itemize}

%\vspace{2pt}
%\noindent
%These metrics provide complementary views of generated implementations but should not be computed over the same set of solutions.

\subsubsection{Structural diversity} 
\label{sec:metrics-structural}

We compute the average \textit{CodeBLEU} similarity between pairs of solutions for each problem~\cite{ren2020codebleu}. CodeBLEU combines four complementary components: n-gram and weighted n-gram similarity (assigns greater importance to syntactically meaningful tokens), \gls{AST} similarity, and dataflow similarity. By incorporating lexical, syntactic, and semantic information, CodeBLEU provides a comprehensive measure of structural similarity between solutions. We use the component weights (0.1, 0.1, 0.4, 0.4) as recommended by the metric's authors~\cite{ren2020codebleu}. As CodeBLEU includes n-gram-based components that are inherently language-specific, we restrict our comparisons to implementations written in the same language.

\subsubsection{Behavioral diversity} 
\label{sec:behavior}

We consider two solutions $i$ and $j$ to be failure-independent if, for every test input $t$, the event that $i$ fails $t$ is statistically independent of the event that $j$ fails $t$: $P(\mathrm{fail}_i(t) \wedge \mathrm{fail}_j(t)) = P(\mathrm{fail}_i(t)) \cdot P(\mathrm{fail}_j(t))$. Under this condition, the expected number of shared failures over a test suite of size $T$ equals $F_i F_j / T$, with $F_i$ and $F_j$ corresponding to the number of tests the solutions fail.

To quantify \textbf{behavioral diversity}, we analyze the \textit{overlap between failure patterns across solutions}. Our approach is inspired by Knight and Leveson’s analysis of failure independence in \gls{NVP}~\cite{knight2012Experimental}, which compares observed joint failures with the frequency expected under a probabilistic model based on global failure rates. In contrast, we use a pairwise formulation that evaluates relationships between individual implementations while conditioning on the number of failures produced by each. This provides a more fine-grained characterization and controls for differences in solution quality.

We compare the observed number of co-failures $O_{ij}$ with the expected number under independence. Under the null hypothesis of independent failures, conditioned on their counts, failed tests are assumed to be randomly distributed among the $T$ tests. If implementation $i$ fails $F_i$ tests and implementation $j$ fails $F_j$ tests, the number of shared failures corresponds to the overlap between two subsets of size $F_i$ and $F_j$ drawn without replacement from $T$. Under this assumption, the expected overlap ($E_{ij}$) follows a hypergeometric distribution~\cite{johnson2005univariate}:

{
\setlength{\abovedisplayskip}{1pt}
\setlength{\belowdisplayskip}{1pt}
\setlength{\belowdisplayshortskip}{1pt}
\setlength{\belowdisplayshortskip}{1pt}

\vspace{-5pt}
\begin{equation}
   E_{ij} = \frac{F_i F_j}{T}
\end{equation}
}

To quantify deviations from this baseline, we compute:

{
\setlength{\abovedisplayskip}{1pt}
\setlength{\belowdisplayskip}{1pt}
\setlength{\belowdisplayshortskip}{1pt}
\setlength{\belowdisplayshortskip}{1pt}

\vspace{-8pt}
\begin{equation}
    Z_{ij} = \frac{O_{ij} - E_{ij}}{\sqrt{\mathrm{Var}_{ij}}}, \quad
    \mathrm{Var}_{ij} = \frac{F_i F_j (T - F_i)(T - F_j)}{T^2 (T - 1)}
\end{equation}
}

This statistic measures how the observed co-occurrence of failures deviates from independence. Positive values indicate failures co-occur more often than expected (correlation), values near zero indicate independence, and negative values indicate complementary behavior (failures on different tests).

This formulation controls for differences in implementation quality, as the expected overlap depends on each implementation’s failure count. Consequently, implementations that fail many tests are not considered correlated unless they fail the same tests more often than expected, isolating failure correlation from overall correctness.

The hypergeometric baseline assumes test exchangeability, i.e., failures are equally likely across tests under independence. In practice, tests vary in difficulty, and some are more prone to failure. To account for this, we analyze two categories separately: standard functional tests and edge-case tests. As discussed next, the former captures typical inputs, while the latter targets boundary conditions. This separation reduces the risk that difficult edge cases dominate the metric, enabling a clearer distinction between general failure correlation and correlations specific to rare scenarios.

\subsubsection{N-version reliability}
\label{sec:metrics-reliability}

To translate behavioral diversity findings into a concrete reliability outcome, we simulate N-version majority voting for $N \in \{3, 5\}$ implementations. Given a set of $M$ implementations for a problem, we sample subsets $S$ of size $N$ and apply a majority decision rule: for each test case, the $N$-version system built from $S$ passes if at least $\lfloor N/2 \rfloor + 1$ of its $N$ implementations pass it. We define $\mathrm{correct}(S) = 1$ if the system passes every test case for that problem, and $0$ otherwise; for $N=1$, $\mathrm{correct}(\{i\})$ reduces to whether implementation $i$ alone passes all tests. \textbf{Absolute reliability} is the fraction of (problem, subset) combinations with $\mathrm{correct}(S)=1$, averaged across all problems.

Absolute reliability is necessary but not sufficient to assess the benefit of voting. It reflects both the quality of the individual implementations and the amount of useful failure diversity among them. A set of weak implementations may have low absolute reliability even if their failures are relatively independent, while a set of strong but highly correlated implementations may have high absolute reliability while gaining little from redundancy. 
%\jrc{this past sentence is difficult to read} 
We therefore complement absolute reliability with \textbf{redundancy effectiveness}: the fraction of the reliability gain expected under independent failures that is actually realized by majority voting. Under the same test-exchangeability assumption used for $Z_{ij}$, the empirical pass rate of implementation $i$ on a single test is $1 - F_i/T$. We apply this to $N$ implementations jointly to measure the probability that a strict majority passes a single test under independence:

{
\setlength{\abovedisplayskip}{0pt}
\setlength{\belowdisplayskip}{0pt}
\setlength{\belowdisplayshortskip}{0pt}
\setlength{\abovedisplayshortskip}{0pt}

\begin{equation}
    P_{\mathrm{maj}}(S) = \sum_{\substack{K \subseteq S \\ |K| > N/2}} \prod_{i \in K} \!\left(1 - \tfrac{F_i}{T}\right) \prod_{i \in S \setminus K} \tfrac{F_i}{T},
\end{equation}

}

where $K$ ranges over all subsets of $S$ in which a majority passes. Under the same test-exchangeability assumption, every test is treated as an independent trial with this majority-pass probability $P_{\mathrm{maj}}(S)$. The system is fully correct only if the majority passes \emph{every} test, so the reliability $S$ would achieve under independence is $P_{\mathrm{maj}}(S)$ multiplied across all $T$ tests. Then, let $\overline{\mathrm{correct}}(S) = \frac{1}{|S|}\sum_{i \in S}\mathrm{correct}(\{i\})$ be the average individual reliability of $S$'s members, $\mathrm{headroom}(S) = R_{\mathrm{indep}}(S) - \overline{\mathrm{correct}}(S)$ the best voting could add under independence, and $\mathrm{gain}(S) = \mathrm{correct}(S) - \overline{\mathrm{correct}}(S)$ what it actually added. Redundancy effectiveness is then

\begin{equation}
    \mathrm{eff}(S) \;=\; \frac{\mathrm{gain}(S)}{\mathrm{headroom}(S)},
\end{equation}

This metric is undefined when $\mathrm{headroom}(S) \approx 0$, so we report it aggregated as $\sum \mathrm{gain}(S) / \sum \mathrm{headroom}(S)$ over a group of sets rather than as an average of per-set ratios. An effectiveness near $1$ means voting captured nearly all of the available gain; near $0$ means almost none of it was realized.

\subsection{Dataset}

A crucial component of the methodology is the dataset, defining the data used for generation and evaluation. It must include a wide range of problems with varying complexity, each accompanied by a problem statement and a test suite.

\begin{itemize}[leftmargin=*]
    \item \textbf{Problem Statements.} Provided as input to the \glspl{LLM}, they define the functional requirements. Each should clearly specify expected behavior, input/output formats, and constraints, ensuring consistent generation conditions.

    \item \textbf{Test Suites.} Define the conditions for executing and comparing the solutions. To ensure meaningful assessment, it should cover the input space comprehensively, including typical and edge cases, exercising different execution paths.

\end{itemize}

%\vspace{2pt}
\noindent
The dataset used in our experiments is based on the \texttt{PROBE} benchmark (\href{https://huggingface.co/datasets/OSS-forge/PROBE}{huggingface.co/datasets/OSS-forge/PROBE}), containing 1,651 contest-style problems from IBM's Project CodeNet dataset~\cite{puri2021codenet}, each with a problem statement, human-written reference solutions, at least four unit tests, and 150 \gls{LLM}-generated solutions (six models, five programming languages, and five samples per model–language pair). We selected this dataset for its scale, diversity, multiple implementations, and reference solutions, enabling correctness validation and test suite augmentation. In addition, its contest-style problems have precise, unambiguous specifications and a single correct behavior, enabling meaningful behavioral comparisons. While \gls{NVP} is ultimately applied to larger software systems, it operates at the level of individual components with well-defined functionality, making these self-contained programming tasks a suitable proxy for studying behavioral diversity.

In its current form, the \texttt{PROBE} dataset is not suitable for diversity and correlated failure analysis. Not all problems are appropriate: some are too simple, with most LLM-generated implementations correct and structurally similar, yielding few failures and limited variation in fault patterns; others are too difficult, with most implementations failing, reflecting model incapacity rather than meaningful differences in strategies. Furthermore, the test suites, scraped from contest platforms and extended with generated inputs and reference solutions, are sufficient for basic correctness evaluation but lack the coverage needed to characterize behavioral differences and correlated failures, which often emerge under specific inputs, boundary cases, or less common execution paths. 

\begin{figure}[t]
    \centering
    \includegraphics[width=0.9\linewidth, trim=0pt 8pt 0pt 8pt, clip]{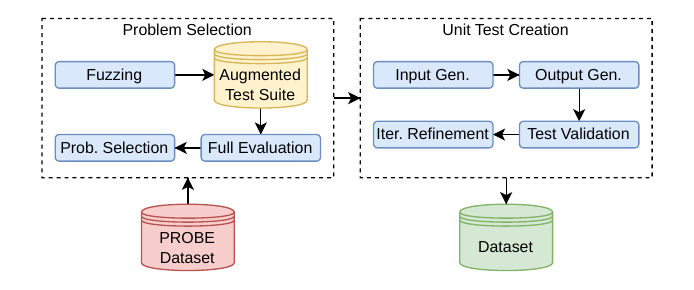}
    \caption{Dataset construction steps}
    \label{fig:dataAug}
\end{figure}

To address these limitations, we curated the dataset following the process illustrated in Figure~\ref{fig:dataAug}, consisting of two steps: (i) selecting a subset of problems suitable for our analysis and (ii) constructing comprehensive unit test suites.

\subsubsection{Problem Selection} The first step filters out problems that are too simple or too difficult, ensuring the selected set supports meaningful analysis of failure diversity. Since the original test suites lack sufficient coverage, we first augment them using coverage-guided \textbf{fuzzing} with AFL++~\cite{fioraldi2020AFL++}, mutating existing test cases and retaining those exploring new execution paths in the reference implementations, until at least 20 unit tests are obtained per problem. Although this does not guarantee systematic coverage, it increases input diversity and execution coverage, providing a stronger basis for evaluation.
        
We then conduct a \textbf{full evaluation} of all LLM-generated implementations in PROBE using the augmented test suites and apply our \textbf{problem selection} criteria. We exclude problems where fewer than 50\% of implementations pass at least 33\% of the tests (i.e., seven), or more than 50\% pass at least 95\% (i.e., fully correct or missing at most one test), ensuring problems are neither trivial nor intractable. We also exclude problems whose reference solutions average fewer than 50 lines, maintaining sufficient structural complexity to support diverse fault types. These thresholds are a practical compromise between behavioral diversity and evaluation tractability. A sensitivity analysis varying each threshold showed smooth, non-abrupt changes in the number of selected problems, indicating our conclusions are unlikely to be significantly impacted by the choice of thresholds (full results at \href{https://tinyurl.com/CodeDiversity}{tinyurl.com/CodeDiversity}). Importantly, the resulting dataset remains large, providing a substantial basis for analysis: the criteria filtered out 824 problems as too difficult, 342 as too easy, and 255 due to size. Six problems were excluded for allowing multiple valid outputs, making correctness difficult to assess, yielding a final set of 224 problems. Although our study uses \texttt{PROBE}, the methodology is not tied to this benchmark. It can be applied to other datasets, including newly generated problems, more complex tasks, or domain-specific workloads, provided they offer executable specifications and reliable test oracles. The same suitability checks are still required: problems should avoid cases where nearly all implementations pass or fail, and test suites must be strong enough to expose behavioral differences beyond basic correctness.

\subsubsection{Unit Test Creation} For the selected problems, we construct refined test suites instead of reusing fuzzing-generated tests, which may violate problem constraints.

Starting with \textbf{Input Generation}, inputs are generated from problem description using the multi-agent approach proposed in~\cite{wang2025codecontests+}: one agent produces scripts generating diverse test cases (random, corner, adversarial), while another verifies compliance with problem constraints. Both agents use GPT-5-mini, excluded from the evaluated models to avoid bias. We generate ten non-edge test cases per problem, plus edge cases covering boundary values and special input structures (e.g., fully connected graphs). We favor this approach over mutation-based methods, which may violate constraints, manual construction, which is costly and hard to scale, and direct LLM-based generation, which does not guarantee constraint satisfaction; our methodology is not tied to this specific technique, however, and only requires a sufficiently diverse, valid unit test set.

In the \textbf{Output Generation} step, expected outputs are produced using human-written reference solutions. To ensure reliability, \textbf{Test Validation} evaluates each test suite against three randomly selected reference implementations, measuring line, branch, and condition coverage, as well as mutation scores via \texttt{Cosmic Ray}~\cite{cosmicray}. Three implementations balance bias avoidance with the feasibility of manually inspecting uncovered regions and surviving mutants to determine whether they result from missing input scenarios or non-impactful code (e.g., redundant checks, equivalent mutations). Finally, in the \textbf{Iterative Refinement} step, suites failing to meet coverage or mutation thresholds undergo targeted manual refinement (rarely needed) until full branch coverage and high mutation scores are achieved across all reference implementations.

After unit test creation, we obtain 10 non-edge tests per problem, along with a variable number of edge cases (20.6 on average). Test suites achieve, on average, 96.7\% line, 94.8\% branch, and 87.6\% condition coverage, with an 86.1\% mutation score. Perfect coverage and mutation scores are not always feasible, as some code is irrelevant to the specification or unreachable with valid inputs, and some mutants survive due to equivalent changes or unobservable effects.

\subsection{Prompts}

Prompting plays a central role in controlling how implementations are generated. To ensure both consistency and controlled variation, we define two key components:

\begin{itemize}[leftmargin=*]

    \item \textbf{Prompting Strategies.} The prompting strategies must be defined to ensure that implementations are produced under controlled and reproducible conditions, while allowing systematic variation in how models approach the problem. 

    \item \textbf{Prompting Templates.} Once the strategies are defined, prompt templates should be specified to ensure their consistent application across models and problems.
    
\end{itemize}

%\vspace{2pt}
\noindent
We evaluate three prompting strategies: Baseline, \gls{CoT}, and Out-of-the-box prompting. We exclude In-Context Learning, as prior work reports negligible effects on functional correctness and code structure~\cite{nogueira2025beyond}. Baseline prompting consists of a direct instruction without guidance. In \gls{CoT} prompting, the model is encouraged to reason step-by-step before generating a solution. Out-of-the-box prompting, instructs the model to produce unconventional or creative solutions, promoting deviation from standard approaches.

\subsection{Procedure}
\label{sec:methodology-procedure}

This section describes the five-phase procedure for generating, evaluating, and analyzing implementations:

\begin{enumerate}[leftmargin=*]

    \item \textbf{Setup:} Includes selecting models and hyperparameters, programming languages, the number of samples to generate for each model–language–prompting strategy, and the criteria used to filter out incorrect implementations from the structural and behavioral diversity analysis.

    \item \textbf{Code Generation:} The models should then be prompted to generate implementations for each problem, according to the selected prompting strategies and templates.

    \item \textbf{Solution Filtering:} After generating all solutions, those that fail basic functional correctness checks should be excluded, as they reflect generation failures rather than different problem-solving strategies and would inflate diversity metrics with noise rather than meaningful variation.

    \item \textbf{Code Evaluation:} The remaning solutions should then be evaluated against the test suite and the results should be processed to obtain the diversity metrics.

    \item \textbf{Fault Analysis:} Identifies and compares the faults responsible for incorrect behavior by analyzing failing implementations Implementations with the same fault are considered failure-correlated, while those failing for different reasons contribute to fault diversity. This provides a more fine-grained view of correlated failures beyond test outcomes.

\end{enumerate}

\noindent

Table~\ref{tab:exp-settings} presents the experimental settings for this study. We consider twelve models selected to span a diverse set of characteristics. In particular, the selection includes proprietary and open-source models, general-purpose and code-specialized systems, and both fully dense and \gls{MoE} architectures. It also covers a range of release periods, from more recent models (e.g., Deepseek-3.2) to slightly older ones (e.g., Qwen2.5-Coder). This diversity is intended to maximize variation in coding styles, reasoning behaviors, and failure modes, providing a comprehensive basis for analysis.

%\textcolor{red}{algum deste conteúdo parece ser mais da parte experimental do que da metodologia. Por exemplo, a escolha de temperatura e número de testes. Também a análise nos penúltimos e antepenúltimos parágrafos é mais de resultados… será de manter aqui?}

The temperature was fixed at 0.6, as prior work shows this achieves strong performance when generating five samples per problem~\cite{chen2021evaluatinglargelanguagemodels}. The remaining hyperparameters were left at default values, standard practice when adjusting temperature. Five programming languages with different characteristics were considered: Python, C++, Java, C, and Rust. Each model generated five samples per language and prompting strategy. 

For the filtering step, we set apart three basic unit tests and discarded all solutions that did not pass these. This choice mirrors the usual \gls{NVP} setting, where each candidate version is first validated against a small test battery before being admitted to the ensemble~\cite{knight2012Experimental}. To check the results were not an artifact of this threshold, we repeated the analysis with several alternative rules: varying the number of required tests from one to three, and replacing the held-out tests with a percentage-based criterion requiring solutions to pass between 10\% and 90\% of all tests. These changed the absolute number of retained solutions, but not the main conclusions, both in terms of overall failure correlations and reliability gains, as well as the relative effects of model, language, and prompting strategy.

\begin{table}[t]
    \centering
    \small
    \caption{Experimental Settings}
    \label{tab:exp-settings}
    \renewcommand{\arraystretch}{1.2}
    \begin{tabular}{r|l}
            \textbf{LLMs} & \makecell[l]{GPT-4.1-mini, Gemini-2.0-flash, Claude-3.0-Haiku, \\ Claude-4.5-Haiku, Qwen-2.5 (14b), Gemma-3 (12b), \\ Ministral-3 (14b), GPT-OSS (120b), Deepseek-v3.2, \\ Qwen-2.5-Coder (7b/14b), Deepseek-Coder-v2 (16b)} \\
            \textbf{Hyper.} & \makecell[l]{Temp. = 0.6, Top-k = 50, Top-p = 1} \\
            \textbf{Langs.} & Python, C++, Java, C, Rust \\
            %\textbf{Samples} & 5
    \end{tabular}
\end{table}

Fault analysis is conducted by inspecting test failures to identify root causes (e.g., boundary errors). We adopt a manual approach because automated fault identification is limited: test outcomes do not uniquely determine the underlying defect. Distinct faults can produce identical behavior, while similar faults may manifest differently depending on the input. Thus, manual inspection remains the most reliable way to determine whether failures share the same root cause.

\section{Results and Discussion}
\label{sec:results}

Our methodology evaluates generated code along three complementary dimensions: structural and behavioral similarity, and reliability gains, complemented by manual analysis. Specifically, we investigate: \textbf{RQ1}, how structurally diverse the generated solutions are; \textbf{RQ2}, whether independently generated implementations fail independently or exhibit correlated failures; \textbf{RQ3}, the reliability gains achieved by N-version majority voting; and \textbf{RQ4}, the extent to which failures across implementations originate from the same underlying faults. Due to space constraints, we present only the most relevant results here; the complete results are available online.

All metrics are computed over the filtered solution set. Overall, 40.8\% of the solutions failed this filtering step and were discarded; 40.2\% were fully correct, passing every test, and 19.0\% were partially correct, passing the filtering tests but not all others. Among the retained solutions, the distribution of fully correct implementations varies primarily by model. The strongest models (GPT-OSS, 80.0\%; DeepSeek-v3.2, 78.9\%; GPT-4.1-mini, 71.2\%) produce substantially more fully correct solutions than the weakest models (below 30\%), with Claude-4.5-Haiku (60.7\%) and Gemini-2.0-flash (44.6\%) lying in between. Programming language has a smaller effect, with Java (45.0\%), Python (42.6\%), and C++ (42.3\%) achieving higher correctness than C (34.8\%) and Rust (36.2\%). Differences across prompting strategies are comparatively minor, with \gls{CoT} (42.8\%), baseline (40.0\%), and out-of-the-box prompting (37.8\%) yielding similar correctness.

\subsection{RQ1: Structural Diversity}

\begin{figure}[t]
    \centering
    \includegraphics[width=1.0\linewidth, trim=5 8 5 0, clip]{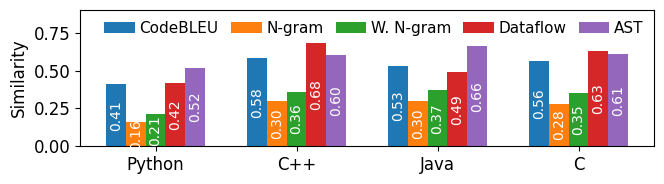}
    \caption{Average CodeBLEU similarity across languages.}
    \label{fig:summary-struc-sim}
\end{figure}

To answer RQ1, we analyze pairwise structural similarity using CodeBLEU. Figure~\ref{fig:summary-struc-sim} shows the average CodeBLEU across languages, which remains relatively high for compiled languages, with mean values of 0.58 for C++, 0.56 for C, and 0.53 for Java. Python exhibits substantially lower similarity (0.41).
This difference is driven by the n-gram and weighted n-gram components, suggesting Python solutions differ more at the surface level, likely due to its expressive syntax and the availability of high-level abstractions~\cite{python_docs}. In contrast, AST and dataflow similarities are closer across languages, indicating that implementations often preserve similar underlying computational structures despite syntactic variation.

\begin{figure}[t]
    \centering
    \includegraphics[width=1.0\linewidth, trim=5 5 5 0, clip]{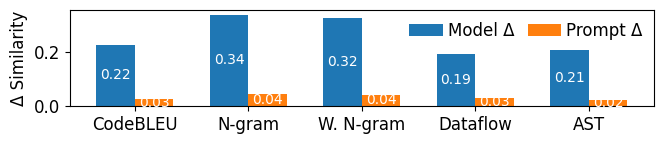}
    \caption{Impact of model/prompting on CodeBLEU score.}
    \label{fig:struc-impact}
\end{figure}

\textbf{Model and Prompting Impact.}
Figure~\ref{fig:struc-impact} compares similarity between implementations generated under the same or different conditions. Implementations from the same model are substantially more similar to each other than those from different models across all languages. For example, in C++, similarity increases from 0.55 (different models) to 0.76 (same model), with similar gaps observed in other languages. This indicates models tend to follow consistent structural patterns when generating multiple solutions for the same task.
This effect is most pronounced in the n-gram and weighted n-gram components, suggesting within-model similarity is largely driven by recurring surface-level patterns (e.g., common templates). Prompting strategies have a much smaller effect. Across all languages, using the same prompting strategy only slightly increases similarity (0.02--0.03), indicating prompting has limited influence compared to model choice.

\textbf{Pairwise Analysis.}
To further analyze the role of model choice, we examine average pairwise similarity between implementations generated by different model pairs. Results show clear differences in within-model consistency (i.e., average pairwise CodeBLEU among samples from the same model): some produce highly similar implementations (e.g., Gemma3, 0.88), while others generate more diverse outputs (e.g., Qwen2.5, 0.67). This indicates models differ not only in functional correctness but also in the rigidity of their generation patterns. These observations are based on the full pairwise similarity analysis in the released dataset. Unlike functional correctness, we do not observe a clear relationship between model size and structural similarity, suggesting that consistency and correctness capture different aspects of model behavior. However, stronger models tend to be less similar to other models overall, while weaker models are more alike. A plausible explanation is that stronger models better capture problem constraints and edge cases, leading to more precise and less common implementations, whereas weaker models rely on generic, shared solution patterns.

A similar pairwise analysis for prompting strategies shows that structural similarity remains highly consistent across different prompts. Across all pairs, similarity values lie within a narrow range (e.g., 0.56-0.60 in C++), and within-prompt similarity is only marginally higher than cross-prompt similarity.

\textbf{Practical implications.}
These results suggest that achieving meaningful structural diversity in practice requires combining heterogeneous models, as varying prompting strategies alone are insufficient. However, the observed diversity is largely concentrated at the surface level, indicating that combining models may not yield fundamentally different solution strategies. Consequently, structural diversity alone may be an unreliable measure for deeper implementation diversity, motivating the need for complementary behavioral or fault-level analyses. 

\begin{tcolorbox}
\textbf{Main takeaways (RQ1).}
Generated implementations exhibit some structural diversity. Outputs from the same model are highly similar; diversity increases across models. Prompting has minimal impact. Model differences shape this: stronger models produce more distinct implementations, whereas weaker ones produce more generic solutions. Diversity is more pronounced at the surface level, while deeper properties, e.g. control flow and data dependencies, vary less.
\end{tcolorbox}

\subsection{RQ2: Behavioral Diversity}

To assess behavioral diversity, we analyze failure pattern similarity using the score introduced in Section~\ref{sec:methodology-metrics}, where 0 indicates overlap expected under independence, negative values indicate less correlated failures, and positive values indicate more correlated failures. We first present results for non-edge-case inputs, then revisit edge-case inputs.

\begin{figure}[t]
    \centering

    % Top row (2 images side by side)
    \begin{subfigure}{0.57\linewidth}
        \centering
        \includegraphics[width=\linewidth,  trim=5 8 5 0, clip]{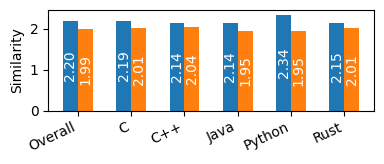}
        \caption{Languages.}
    \end{subfigure}
    \hfill
    \begin{subfigure}{0.4\linewidth}
        \centering
        \includegraphics[width=\linewidth,  trim=5 8 5 0, clip]{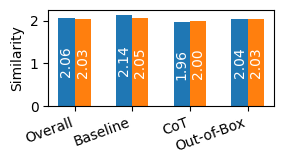}
        \caption{Prompts.}
    \end{subfigure}

    \vspace{0.57em}

    % Bottom row (single centered image)
    \begin{subfigure}{1.0\linewidth}
        \centering
        \includegraphics[width=\linewidth, trim=5 10 5 0, clip]{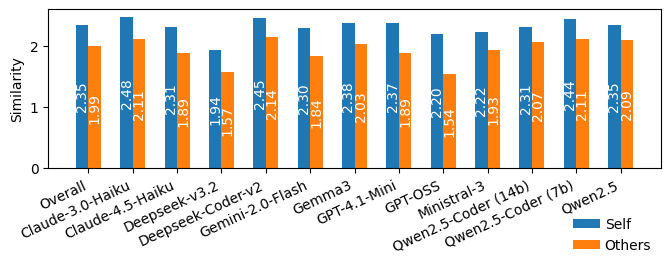}
        \caption{Models.}
        \vspace{2pt}
    \end{subfigure}
    \caption{Behavioral Sim. within the same/different entities}
    \label{fig:summary_behavior_similarity}
\end{figure}

Figure~\ref{fig:summary_behavior_similarity} shows the average pairwise similarity between implementations generated under the same or different conditions. Across all implementations, the mean similarity is 2.04, indicating a clear deviation from independence. Therefore, failures are not randomly distributed, co-occurring on the same inputs. To assess significance, we compute, for each problem, the proportion of implementation pairs with a statistically significant positive deviation ($\alpha = 0.05$), and average across problems. On average, 65.9\% of pairs significantly deviate from independence, indicating this effect is systematic, rather than driven by a small subset of cases. Due to space constraints, we report this statistical analysis only for the overall results; applying it to the remaining breakdowns yields trends consistent with those observed from the metric values alone.

\textbf{Language Impact.}
Across languages, differences are noticeable, though still relatively moderate. The gap between within-language and cross-language similarity is consistent, indicating that implementations written in the same language tend to fail more on similar inputs. Among languages, Python stands out, exhibiting the highest within-language similarity (2.34) while maintaining lower similarity with other languages (e.g., 1.88 with Java). This suggests that Python implementations are more internally aligned in their failure patterns, which can be explained by relying more heavily on high-level libraries and abstractions, reducing the space of possible failure modes and leading to more concentrated, similar errors.

\textbf{Prompting impact.}
Prompting strategies have minimal effect, as differences between within- and cross-prompt similarity are negligible (-0.03), and no prompting pair exhibits substantially higher/lower similarity. This is consistent with the limited structural variation observed in RQ2.

\textbf{Model impact.}
Model choice has a strong impact on behavioral similarity. The gap between within- and cross-model similarity is more pronounced (-0.36), indicating solutions from the same model tend to fail on the same inputs. This effect correlates with model performance. Stronger models show larger separation between within- and cross-model similarity and lower similarity with other models, indicating more distinct failure patterns. However, within-model similarity remains high, meaning that although their failures differ from other models, they remain strongly correlated across samples. Section~\ref{sec:fault-analysis} suggests this is because stronger models avoid simpler errors but still exhibit consistent internal error patterns.

\textbf{Pairwise analysis.}
Pairwise breakdowns confirm the trends above. Among languages, compiled-language pairs (e.g., C++/Rust: 2.08, C/C++: 2.09) are more correlated than those involving Python, while all prompting pairs lie in a narrow range (1.98 to 2.08). Model pairs show the widest spread: the least correlated pairs (GPT-OSS/Gemini-2.0-flash: 0.80) contrasts sharply with the most correlated weak-model pairs (Claude-3.0-Haiku/DeepSeek-Coder-v2: 2.40), confirming that stronger models exhibit more distinct failure modes.

\textbf{Edge-Case analysis.} 
When restricting the analysis to edge-case inputs, the overall behavioral similarity increases to 3.39, indicating a stronger deviation from independence. Correspondingly, the proportion of pairs exhibiting statistically significant positive deviation rises to 71.78\%. This shows that failures become more aligned under more challenging conditions, with implementations more consistently failing on the same inputs (e.g., inputs at the boundaries of the problem constraints, such as maximum input size, or max/min values).

Across languages and prompting strategies, trends observed for non-edge cases persist with minor differences. However, model choice has a stronger effect: the gap between within- and cross-model similarity increases substantially (overall gap of -0.78), especially for stronger models such as GPT-OSS (-1.47) and GPT-4.1-mini (-1.08). This suggests solutions from the same model become more aligned in their failure patterns on difficult inputs, likely because edge cases expose a limited set of challenging conditions. Pairwise analysis confirms this: although similarities increase overall, indicating stronger correlation, the relative structure remains unchanged (stronger models still show lower similarity with others).

\textbf{Practical implications.}
These results suggest that increasing structural diversity alone is insufficient to achieve failure independence. Even when combining different models, implementations frequently fail on the same inputs, particularly in challenging scenarios, limiting the effectiveness of \gls{NVP} and highlighting the need for approaches that explicitly target behavioral diversity rather than relying solely on model or prompt variation. However, the observation that certain model pairs exhibit substantially lower behavioral similarity suggests that carefully selecting and combining complementary models may help reduce failure correlation, motivating strategies for constructing such ensembles that explicitly optimize for behavioral diversity rather than relying on ad hoc combinations.

\begin{tcolorbox}
\textbf{Main takeaways (RQ2).}
Failures are strongly correlated across implementations, indicating clear deviation from independence. Model choice has the strongest effect: more complex models produce more distinct failure patterns, yet failures remain highly correlated within the same model. Language introduces some variation, while prompting has negligible impact. Correlation is further amplified on edge-case inputs, where failures become even more aligned.
\end{tcolorbox}

\subsection{RQ3: Reliability Gains}

To quantify the practical impact of failure correlation, we measure N-version reliability under majority voting for $N \in \{3, 5\}$ using the metrics in Section~\ref{sec:metrics-reliability}. Table~\ref{tab:reliability-overall} presents overall absolute reliability and redundancy effectiveness for $N \in \{3, 5\}$, with the homogeneous (all $N$ versions from the same language/model/prompt) and heterogeneous (versions drawn from at least two distinct values) breakdown for each dimension. Overall reliability improves from 0.88 (average reliability of all solutions) to 0.90 at N=3 and 0.91 at N=5, a gain of only 0.03, while redundancy effectiveness stays low and roughly flat (0.43 at N=3, 0.44 at N=5): majority voting never captures more than half of the reliability gain that would have been available under independence, and this fraction does not improve as more versions are added.

\begin{table}[t]
    \centering
    \footnotesize
    \caption{N-version reliability/Redundancy effectiveness} 
    \label{tab:reliability-overall}
    \setlength{\tabcolsep}{2.2pt}
    \renewcommand{\arraystretch}{1.1}
    \begin{tabular}{c c cc cc cc}
    \toprule
    & & \multicolumn{2}{c}{Language} & \multicolumn{2}{c}{Model} & \multicolumn{2}{c}{Prompting} \\
    $N$ & Overall & Hom & Het & Hom & Het & Hom & Het \\
    \midrule
    1 & 0.88/-- & -- & -- & -- & -- & -- & --   \\
    3 & 0.90/0.43 & 0.90/0.37 & 0.90/0.44 & 0.92/0.27 & 0.90/0.44 & 0.90/0.41 & 0.90/0.44 \\
    5 & 0.91/0.44 & 0.92/0.37 & 0.91/0.44 & 0.94/0.24 & 0.91/0.44 & 0.90/0.42 & 0.91/0.44 \\
    \bottomrule
    \end{tabular}
\end{table}

\textbf{Language, Model, and Prompting Impact.}
Table~\ref{tab:reliability-overall} shows that redundancy effectiveness consistently favors heterogeneous combinations. The largest improvement comes from combining different models (0.27 vs. 0.44 at N=3; 0.24 vs. 0.44 at N=5), followed by programming language (0.37 vs. 0.44), while prompting has only a small effect (0.41 vs. 0.44 and 0.42 vs. 0.44). These trends mirror the behavioral-diversity results from RQ2, indicating that the dimensions producing less correlated failures also produce the greatest redundancy benefit. This trend is consistent across individual models: at N=3, all twelve models achieve higher redundancy effectiveness when combined with different models than with additional instances of themselves. Per-problem paired Wilcoxon tests confirm that model diversity is strongly significant at N=3 ($p<0.001$); language and prompting are also significant ($p=0.007$ and $p=0.043$) but with much smaller effects.

Note that, while absolute reliability is reported for completeness, the hom-versus-het comparison should be interpreted primarily through redundancy effectiveness. Unlike absolute reliability, redundancy effectiveness controls for differences in the quality of the underlying implementations and therefore isolates the benefit obtained from diversity. Absolute reliability is influenced by the filtering step, which causes strong models to contribute disproportionately many homogeneous combinations, artificially inflating their average reliability.

\textbf{Pairwise Analysis.}
The pairwise results mirror RQ2. Language and prompting pairs have little impact, with effectiveness in a narrow range (0.41 to 0.45 and 0.42 to 0.46 at $N=3$, respectively). Model pairs vary substantially, and their effectiveness closely follows the behavioral-correlation ranking from RQ2: the least correlated pair (GPT-OSS/Gemini-2.0-flash, $Z=0.80$) achieves the highest effectiveness (0.59/$N=3$, 0.69/$N=5$), whereas the most correlated pairs rank near the bottom, reinforcing the weak-versus-strong model split observed earlier. Pairs of weak models remain unreliable and can even degrade with additional versions (Ministral-3-14b/Qwen2.5-Coder-7b is the only pair with negative effectiveness at $N=5$), whereas pairs of strong models achieve the highest absolute reliability but only moderate effectiveness. Notably, even the strongest pair (GPT-4.1-mini/GPT-OSS, 0.97 reliability) does not surpass GPT-OSS alone (0.98), indicating that failure correlation limits the benefit of redundancy.

\textbf{Practical implications.}
These results suggest that the cost-benefit case for redundancy must be evaluated carefully: adding more versions yields diminishing and modest returns, capturing less than half of what independence would allow. Second, if redundancy is pursued, model diversity should be the primary lever, as language and prompting variations contribute comparatively little. Finally, practitioners should benchmark the strongest single model before committing to a multi-version ensemble, since even the best heterogeneous combination fails to surpass it in absolute reliability.

\begin{tcolorbox}
\textbf{Main takeaways (RQ3).}
N-version majority voting yields modest reliability gains, capturing less than half the gain available under independence. Model diversity drives the largest improvements, while language and prompting have little effect. No combination surpasses the strongest single model, confirming that current LLMs fall short of the independence required for effective \gls{NVP}.
\end{tcolorbox}

\subsection{RQ4: Fault Analysis}
\label{sec:fault-analysis}

To better understand the sources of failure correlation, we manually analyzed approximately 650 partially correct implementations from fifteen problems with the highest and lowest average pairwise behavioral similarity (3 highest and 3 lowest for both non-edge and edge tests, plus 3 intermediate). For problems with more than 50 partially failing implementations, we randomly sampled three implementations per distinct failure pattern to keep the analysis feasible.

\textbf{Low behavioral similarity.}
In problems with low behavioral similarity, solutions exhibit varied failure patterns, often failing different subsets of tests. Manual inspection reveals these differences are largely superficial: many implementations share the same underlying issue manifesting differently across inputs. For instance, problem p3685 presents 16 distinct failure patterns among 34 partially correct implementations, yet most stem from the same root cause: reliance on global heuristics rather than reasoning about input structure. The task requires determining connectivity of values on a grid by their spatial arrangement, but several solutions rely only on aggregate quantities (e.g., grid dimensions), failing to distinguish inputs with the same global properties but different layouts. Faults are also highly model-dependent: 11 of the 34 solutions originate from Qwen2.5-Coder:7b and 6 from DeepSeek-Coder-v2, indicating that specific models consistently reproduce the same faulty reasoning patterns.

A superficial analysis of three problems with intermediate behavioral similarity revealed similar trends, with many distinct failure patterns caused by a small set of underlying causes. However, they showed a slightly smaller set of distinct failure patterns (relative to the number of partially correct solutions), explained by fewer variants per cause, leading to higher behavioral similarity scores.

\textbf{High behavioral similarity.}
Problems with high behavioral similarity show a clearer pattern: solutions fail the same way on the same inputs, making shared faults immediately visible. In problem p02818, many implementations adopt an $O(K)$ approach despite an $O(1)$ solution, failing on the same large-input tests. In p04013, stack overflows from large local memory allocations produce identical failure behavior across solutions. Fault concentration in specific models is again evident: 209 of 254 solutions with the timeout issue originate from four models, and the issue is absent in the strongest models, suggesting their lower behavioral similarity may stem from avoiding simple errors that others consistently make.

\textbf{Practical implications.}
These findings suggest that, in the analyzed cases, generating multiple implementations is insufficient to ensure fault diversity, as they rely on similar reasoning patterns. Even when implementations appear behaviorally diverse, they may share the same underlying weaknesses, limiting the effectiveness of redundancy-based approaches and highlighting the need for techniques that promote diversity in reasoning strategies, rather than surface-level details. They also emphasize the importance of systematic methodologies, like the one proposed in this paper, to assess failure independence and uncover hidden correlations.

\begin{tcolorbox}
\textbf{Main takeaways (RQ4).}
Implementations often share the same underlying faults, which manifest as identical failures when behavioral similarity is high and as seemingly different patterns when lower, creating the illusion of diversity. These faults also tend to concentrate within specific models, which consistently reproduce the same faulty reasoning.
\end{tcolorbox}

\begin{comment}
\begin{figure}
    \centering
    \begin{lstlisting}[style=smallc]
...
for (int i = 0; i < N; ++i) {
long long x1 , y1 , x2 , y2 
scanf("%lld %lld %lld %lld", &x1, &y1, &x2, &y2);
} // discards the coordinates
if (R + 1 >= N || C + 1 >= N) {
printf("YES\n");
} else {
printf("NO\n");}
\end{lstlisting}

\begin{lstlisting}[style=smallc]
...
if (N == 1) {
    printf("YES\n");
} else if (N == 2 && R > 1 && C > 1) {
    printf("YES\n");
} else if (N > 2 && R > N - 1 && C > N - 1) {
    printf("YES\n");
} else {
    printf("NO\n");}
\end{lstlisting}
\vspace{-4pt}
    \caption{Incorrect solutions using global heuristics.}
    \label{fig:listings}
\end{figure}
\end{comment}

\section{Threats to Validity}
\label{sec:ttv}

\textbf{Internal Validity.}
A potential threat is data leakage, as the dataset originates from 2021 competitive programming problems and may overlap with the training data of some evaluated models. While we cannot fully rule this out, several lines of evidence suggest that memorization is unlikely to be the primary driver of the observed failure correlation. If leakage were a dominant factor, models with greater training overlap would exhibit notably higher correctness and stronger failure correlation than other models; instead, failure correlation is consistent across all models, regardless of architecture, provider, or training pipeline. Furthermore, if memorization were inflating correlation, the strongest models should exhibit the most correlated failures, as they would be reproducing the same solutions; however, the results show the opposite, with stronger models producing less correlated failure patterns. Additionally, the behavioral analysis is restricted to implementations that fail at least one test, meaning that memorized solutions, which would simply be correct, are excluded and therefore cannot inflate failure correlation. Finally, the manual fault analysis suggests that failures stem from systematic reasoning weaknesses, such as reliance on global heuristics, integer overflows, and asymptotically suboptimal algorithms, rather than from reproducing memorized outputs.

\textbf{External Validity.}
External validity may be limited by the choice of models and problem domains. Regarding models, we do not evaluate the very largest frontier systems, partly due to cost constraints and partly because keeping pace with the latest releases is inherently difficult, with any fixed selection risking becoming outdated within months. We nonetheless evaluate twelve models spanning diverse architectures, sizes, and providers, including recent releases such as GPT-oss-120B, Claude 4.5 Haiku, and DeepSeek-3.2, and the consistent patterns observed across this set suggest the findings are not tied to a specific model family. We also note that practitioners do not always have access to the latest frontier models, so results over this representative set remain practically relevant. Regarding the domain, the study focuses on competitive programming tasks, which differ from real-world software. However, they provide two key advantages: unambiguous specifications and comprehensive, automatable test suites, enabling precise analysis of failure independence. Importantly, these tasks represent a best-case scenario for failure independence: they are self-contained, well-specified problems with a single correct behavior, free of the ambiguity, sprawling dependencies, and evolving requirements that characterize production codebases. If diversity is not observed under these conditions, it is unlikely to emerge in more complex real-world settings. Moreover, while individual competitive-programming-style problems are not representative of full systems, \gls{NVP} is typically applied not to entire systems but to discrete, well-specified, safety-critical components within them, making this granularity a reasonable proxy for the units where such redundancy schemes are actually deployed. Extending this analysis to other domains (e.g., library or systems code) remains future work.

\textbf{Construct Validity.}
We use unit test outcomes as a proxy for correctness and behavioral similarity, which may be limited if test suites do not fully cover the input space. To mitigate this, we construct test suites using automated input generation, validation against multiple human-written reference implementations, and iterative refinement guided by branch, condition, and mutation coverage, achieving high coverage and mutation scores. Since behavioral diversity is measured via pairwise failure comparisons, remaining coverage gaps are unlikely to significantly affect results, as missing cases would impact all implementations similarly.
A further limitation is the assumption of test exchangeability in the hypergeometric baseline, which ignores variation in test difficulty. We partially mitigate this by separating standard and edge-case tests, though variation may still persist within each category.
Finally, test inputs are generated using an LLM-based pipeline, which may introduce stylistic biases that favor certain solution strategies; this is mitigated through diverse test types (random, corner, adversarial) and coverage validation.

\section{Conclusion}
\label{sec:conclusions}

This paper proposes a methodology to evaluate diversity in \gls{LLM}-generated code, combining structural diversity, behavioral diversity, reliability gains, and manual fault analysis. It is instantiated on twelve models, 224 problems, five programming languages, and three prompting strategies.

Results show strong failure correlation across implementations despite structural variation, with models frequently failing on the same inputs. This limits reliability gains under majority voting, which remain below half of the theoretical gains under independence, and no model ensemble surpasses the best single model. Fault analysis confirms that many failures stem from shared underlying issues.

Model choice is the dominant factor: same-model outputs are more similar, more correlated in failure, and yield lower redundancy benefits than cross-model ensembles. Language has a moderate effect, while prompting strategy has negligible influence. Overall, heterogeneous ensembles offer the best, though still limited, improvement in failure independence.

These findings suggest the independence assumption underlying \gls{NVP} does not hold for current \glspl{LLM}, constraining redundancy benefits, and highlight the value of our methodology for assessment of failure independence as \glspl{LLM} evolve.
 
Future work will explore methods to increase failure diversity, including diversity-oriented prompting, fine-tuning, and generation pipelines that explicitly optimize for correctness and behavioral independence, as well as applicability to other multi-version paradigms such as recovery blocks.

\clearpage

\bibliographystyle{IEEEtran}
\bibliography{IEEEabrv,bibliography}

\end{document}